# Economic effects of Chile FTAs and an eventual CTPP accession

Mauricio "Pachá" Vargas Sepulveda

v. 2022-09-28 13:29

**Draft document, NOT peer-reviewed. Please see github.com/pachadotdev/general-equilibrium-ftas-chile for updates and final peer-reviewed publication in the future.**

## Contents

# 1   Summary

In this article, we show the benefits derived from the Chile-USA (in-force Jan, 2004) and Chile-China (in-force Oct, 2006) FTA on GDP consumer and producers to conclude that Chile improved its welfare improved after its subscription. From that point, we extrapolate to show the direct and indirect benefits of CTPP accession.

The models presented here consists are organized in counterfactual simulations consisting in 'what if' scenarios. For in-force FTAs we simulate scenarios where the FTAs were dropped the year after they went fully in-force. For the potential Chile's accession to CTPP, we evaluate it by doing the opposite, by defining a counterfactual scenario where a Chile-CTPP Region would have entered in force one year before the moment of the original signature in Feb, 2016.

In our model, an increase of 0.48% over Chile's GDP comes from the Chile-USA FTA while an increase of 0.22% comes from the Chile-China FTA. An eventual CTPP accession would mean that Chile's GDP increases by 0.99%. In addition, we computed General Equilibrium effects indicating both consumers and producers in member countries of the Chile-USA and Chile-China FTAs face positive effects.

# 2   Introduction

This work proposes an econometric experiment to investigate the general equilibrium effects of FTAs, particularly the effects of the Chile-China and Chile-USA FTAs. The main reference for the steps discussed here is Yotov et al. (2016).

The choice of these agreements is because China and the USA are the No. 1 and No. 2 trading partners with Chile, according to the most recent United Nations trade data for 2021 (United Nations Statistics Division 2022).

Focusing on these FTAs, whose effects are misunderstood in public opinion, primarily due to extreme political positions in Latin America, aims to measure the observed effects of trade on Chile's GDP.

This econometric experiment relies on panel data to identify the effects of FTAs and capture the impact of trade costs by using paired fixed effects (Yotov et al. 2016). The application implements a two-stage procedure to recover missing bilateral trade costs (Anderson and Yotov 2016).

This work solves a General Equilibrium problem, which considers the interaction of different market forces similar to the case of a multiple degrees of freedom spring-mass-damper system that we find in physics textbooks.

The essential ingredient to solve this system is the Poisson Pseudo Maximum Likelihood estimation (Silva and Tenreyro 2006), implemented as a constrained regression with an external trade cost vector, resulting in General Equilibrium PPML.

# 3   Estimating the Impact of FTAs

The analysis begins by specifying the following panel version of the empirical gravity model in order to obtain estimates of bilateral trade costs, including an estimate of the average effects of all FTAs (Yotov et al. 2016):

$$X_{ij,t} = \exp\left[\beta_1 FTA_{ij,t} + \pi_{i,t} + \chi_{j,t} + \mu_{ij}\right] \times \varepsilon_{ij,t}. \tag{1}$$

Where $FTA_{ij,t}$ is an indicator variable equal to one if two countries are members of the same FTA at year $t$, and zero otherwise.

Following the recommendations in Yotov et al. (2016), this specification is estimated with the PPML

estimator using panel data with exporter-time fixed effects ($\pi_{i,t}$) and importer-time fixed effects ($\chi_{j,t}$).

Pair fixed effects ($\mu_{ij}$) are included to alleviate potential endogeneity concerns of the FTA variable and to control for time-invariant trade costs at the bilateral level.

Here we evaluate the hypothetical scenario of removing FTAs after they entered into force in 2004 (USA) and 2006 (China).

## 3.1  Step I: Solve the baseline gravity model

We need to estimate the baseline gravity model (5) to obtain point estimates of the effect of FTAs and the pair fixed effects and constructing the bilateral trade costs matrix required to compute the baseline indexes of interest.

It would be ideal to construct the complete matrix of bilateral trade costs in order to perform sound counterfactual analysis, but it's often not possible to identify the complete set of pair fixed effects estimates in the gravity model when trade flows data are missing (or zero) for a given pair of countries in the period under investigation.

Following Anderson and Yotov (2016), the two-stage procedure is implemented in order to recover the complete set of bilateral trade costs.

We keep the panel dimension of the dataset to identify the effects of FTAs and comprehensively capture the impact of all time-invariant trade costs with the use of pair fixed effects (Vargas Sepulveda 2021).

### 3.1.1  Stage 1: Obtain the estimates of pair fixed effects and the effects of FTAs

For estimation purposes, the panel shall contain the years 1990 to 2006 in intervals of four years.

The first stage consists of estimating equation (5) in order to obtain the estimates of the bilateral fixed effects for country-pairs with non-missing (or non-zero) trade flows:

$$X_{ij,t} = \exp\left[\hat{\beta}_1 \mathsf{FTA}_{ij,t} + \hat{\pi}_{i,t} + \hat{\chi}_{j,t} + \hat{\mu}_{ij}\right]. \qquad (2)$$

### 3.1.2  Stage 2: Regress the estimates of pair fixed effects on gravity variables and country fixed effects

We use the estimates of the pair fixed effects ($\mu_{ij}$) from equation (2) as the dependent variable in a regression where the covariates include the set of standard gravity variables along with importer and exporter fixed effects (Larch and Yotov 2016):

$$t_{ij}^{1-\sigma} = \exp[\mu_{ij}] = \exp\left[\beta_1 \log(\mathsf{DIST}_{ij}) + \beta_2 \mathsf{CNTG}_{ij} + \beta_3 \mathsf{LANG}_{ij} + \beta_3 \mathsf{CLNY}_{ij} + \hat{\pi}_{i,t} + \hat{\chi}_{j,t}\right]. \qquad (3)$$

The predictions from regression (3) are used to fill in the missing trade cost values in order to obtain the complete set of bilateral trade costs $\hat{t}_{ij}^{1-\sigma}$.

Once the full vector of bilateral trade costs is obtained, we include it as a constraint in the baseline gravity specification (2), which will return estimates for the importer and exporter fixed effects. The result is consistent with the trade cost vector and can be used to recover the corresponding values of the multilateral resistances.

A similar constrained estimation procedure should be performed when the trade cost vector is obtained externally, including when it is constructed with a calibration method.

## 3.2    Step II: Define a counterfactual scenario

We need to define the hypothetical removal of the USA and China FTAs with Chile. By re-defining the FTA dummy variable, $FTA_{ij,t}^{CFL}$, as if FTAs were not in place by setting the original FTA indicator variable to be equal to zero for trade between Chile and the USA after 2004 and Chile and China after 2006.

## 3.3    Step III: Solve the counterfactual model

We need to construct the counterfactual indexes of interest in the Conditional GE and the Full Endowment GE scenarios of removing the abovementioned FTAs.

The Conditional GE effects from the removal of the FTAs above are computed by re-estimating the econometric gravity specification (5) for 2004 and 2007, the year of entry into force of the agreements, subject to constraints reflecting the counterfactual scenario:

$$X_{ij} = \exp\left[\hat{\beta}_1 FTA_{ij}^{CFL} + \pi_i^{CFL} + \chi_j^{CFL} + \hat{t}_{ij}^{1-\sigma}\right] \times \varepsilon_{ij}^{CFL}. \tag{4}$$

Equation (3.3) is estimated under the constraints that the Chile-USA and Chile-China FTAs never existed ($FTA_{ij}^{CFL}$) and the coefficient of the FTA dummy and the bilateral fixed effects are equal to their baseline values, $\hat{\beta}_1$ and $\hat{t}_{ij}^{1-\sigma}$ respectively, such that no part of the trade costs besides the FTA dummy changes.

The PPML estimates of the directional effects from equation (3.3) can be used to recover the conditional multilateral resistance indexes $\pi_i^{CFL}$ and $\chi_j^{CFL}$ subject to normalization.

The values of the Full Endowment GE effects of the removal of the FTAs are directly obtained by implementing an iterative procedure, which sequentially allows for endogenous factory-gate prices, followed by income, expenditure and trade to adjust to the counterfactual shock.

Note that solving requires re-estimating the counterfactual model in a loop where $p_i$, $p_j$ and $y$ variables from the Full Endowment change until convergence is reached. The same can be done by explicitly solving the GE non-linear system with any fixed point arithmetic routine.

For this particular case, we define the maximum price difference from iteration $m-1$ to $m$ as $d = |\max(\Delta p_i^m - \Delta p_i^{m-1})|$. When $d \leq 10^{-3}$ and the standard deviation of the price difference vector, $s = \Delta p^m - \Delta p^{m-1}$, is such that $\mathrm{sd}(s) \leq 10^{-3}$, we claim convergence is reached.

In the next sections we report the results of the counterfactual analysis, including the percentage difference between the baseline values and their Full Endowment counterparts of the main variables of interest for each country in the sample.

Despite some specifications and data limitations, the results presented and discussed above are comparable with findings from existing related studies. GEPPML produces a theoretically sound model.

# 4    Simulation results

## 4.1    Dropping the Chile-USA FTA

The estimate of $\beta_1$ in (5) implies that, on average, the Chile-USA FTA has led to $[\exp(\hat{\beta}_1)-1] \times 100 \approx [\exp(0.44) - 1] \times 100 \approx 55$ percent increase in trade among members.

Table 1: Step I-1 model summary for the Chile-USA FTA

| Dependent Variable: | Trade ($X_{ij,t}$) |
|---|---|
| Model: | |
| *Variables* | |
| FTA | 0.4383*** |
| | (0.0987) |
| *Fixed-effects* | |
| Exporter-year ($\pi_{i,t}$) | Yes |
| Importer-year ($\chi_{j,t}$) | Yes |
| Pair ($\mu_{i,j}$) | Yes |
| *Fit statistics* | |
| Observations | 114,745 |
| Squared Correlation | 0.99881 |
| Pseudo R$^2$ | 0.99709 |
| BIC | 2,510,824.4 |

*Clustered standard-errors in parentheses*
*Signif. Codes: ***: 0.01, **: 0.05, *: 0.1*

Table 2: Step I-2 model summary for the Chile-USA FTA

| Dependent Variable: | Trade costs ($\bar{t}_{i,j}$) |
|---|---|
| Model: | (3) |
| *Variables* | |
| Distance ($\log(\text{DIST}_{ij})$) | -1.129*** |
| | (0.1655) |
| Contiguity ($\text{CNTG}_{ij}$) | 0.2049 |
| | (0.6504) |
| Language ($\text{LANG}_{ij}$) | 0.1595 |
| | (0.5241) |
| Colony ($\text{CLNY}_{ij}$) | 1.966*** |
| | (0.5752) |
| *Fixed-effects* | |
| Exporter ($\pi_{i,t}$) | Yes |
| Importer ($\chi_{j,t}$) | Yes |
| *Fit statistics* | |
| Observations | 30,274 |
| Squared Correlation | 0.87682 |
| Pseudo R$^2$ | 0.91504 |
| BIC | 22,558,899.5 |

*Clustered standard-errors in parentheses*
*Signif. Codes: ***: 0.01, **: 0.05, *: 0.1*

Column II of Table 7 reveals that, under the conditional general equilibrium scenario, the member countries of the Chile-USA FTA experience the largest export increase, ranging from 0.23% (USA) to 5.44% (Chile), of their total exports.

Column III of Table 7 reveals that the values of the Full Endowment general equilibrium effects of the FTAs above on exports are qualitatively identical to the corresponding conditional equilibrium effects, despite some quantitative differences.

Column IV of Table 7, the counterfactual analysis, suggests that the Full Endowment welfare effects of the Chile-USA FTA are favourable for its members, ranging from 0.01% (USA) to 0.48% (Chile), and slightly negative or null for non-member countries.

The decomposition of the Full Endowment GE effects reported in columns V, VI, and VII suggests that both consumers and producers in member countries of the FTAs face positive effects with lower inward multilateral resistances (for consumers) and lower outward multilateral resistances, which translate into higher factory-gate prices (for producers) relative to the effects on the consumers in the reference country (i.e., Germany for the current case).

## 4.2    Dropping the Chile-China FTA

The estimate of $\beta_1$ in (5) implies that, on average, the Chile-USA FTA has led to $[\exp(\hat{\beta_1}) - 1] \times 100 \approx [\exp(0.23) - 1] \times 100 \approx 26$ percent increase in trade among members.

Table 3: Step I-1 model summary for the Chile-China FTA

| Dependent Variable:<br>Model: | Trade ($X_{ij,t}$) |
|---|---|
| *Variables* | |
| FTA | 0.2348*** |
| | (0.0493) |
| *Fixed-effects* | |
| Exporter-year ($\pi_{i,t}$) | Yes |
| Importer-year ($\chi_{j,t}$) | Yes |
| Pair ($\mu_{i,j}$) | Yes |
| *Fit statistics* | |
| Observations | 122,043 |
| Squared Correlation | 0.99874 |
| Pseudo $R^2$ | 0.99704 |
| BIC | 2,735,880.9 |

*Clustered standard-errors in parentheses*
*Signif. Codes: ***: 0.01, **: 0.05, *: 0.1*

Table 4: Step I-2 model summary for the Chile-China FTA

| Dependent Variable:<br>Model: | Trade costs ($\tilde{t}_{i,j}$)<br>(3) |
|---|---|
| *Variables* | |
| Distance ($\log(\text{DIST}_{ij})$) | -1.011*** |
| | (0.1445) |
| Contiguity ($\text{CNTG}_{ij}$) | 0.7273** |
| | (0.3315) |
| Language ($\text{LANG}_{ij}$) | 1.574*** |
| | (0.3593) |
| Colony ($\text{CLNY}_{ij}$) | 0.3870 |
| | (0.8475) |
| *Fixed-effects* | |
| Exporter ($\pi_{i,t}$) | Yes |
| Importer ($\chi_{j,t}$) | Yes |
| *Fit statistics* | |
| Observations | 30,943 |
| Squared Correlation | 0.91694 |
| Pseudo R$^2$ | 0.95627 |
| BIC | 10,566,872.2 |

*Clustered standard-errors in parentheses*
*Signif. Codes: ***: 0.01, **: 0.05, *: 0.1*

Column II of Table 8 reveals that, under the conditional general equilibrium scenario, the member countries of the Chile-USA FTA experience the largest export increase, ranging from 0.09% (China) to 2.02% (Chile), of their total exports.

Column III of Table 8 reveals that the values of the Full Endowment general equilibrium effects of the FTAs above on exports are qualitatively identical to the corresponding conditional equilibrium effects, despite some quantitative differences.

Column IV of Table 8, the counterfactual analysis, suggests that the Full Endowment welfare effects of the Chile-USA FTA are favourable for its members, ranging from 0.00% (China) to 0.22% (Chile), and slightly negative or null for non-member countries.

The decomposition of the Full Endowment GE effects reported in columns V, VI, and VII suggests that both consumers and producers in member countries of the FTAs face positive effects with lower inward multilateral resistances (for consumers) and lower outward multilateral resistances, which translate into higher factory-gate prices (for producers) relative to the effects on the consumers in the reference country (i.e., Germany for the current case).

## 4.3   Subscribing the Chile-CTPP Region FTA

The estimate of $\beta_1$ in (5) implies that, on average, the Chile-USA FTA has led to $[\exp(\hat{\beta_1}) - 1] \times 100 \approx$ $[\exp(0.10) - 1] \times 100 \approx 10$ percent increase in trade among members.

Table 5: Step I-1 model summary for the Chile-CTPP Region FTA

| Dependent Variable:<br>Model: | Trade ($X_{ij,t}$) |
|---|---|
| *Variables* | |
| FTA | 0.0995*** |
| | (0.0297) |
| *Fixed-effects* | |
| Exporter-year ($\pi_{i,t}$) | Yes |
| Importer-year ($\chi_{j,t}$) | Yes |
| Pair ($\mu_{i,j}$) | Yes |
| *Fit statistics* | |
| Observations | 133,180 |
| Squared Correlation | 0.99977 |
| Pseudo R$^2$ | 0.99776 |
| BIC | 3,246,192.1 |

*Clustered standard-errors in parentheses*
*Signif. Codes: ***: 0.01, **: 0.05, *: 0.1*

Table 6: Step I-2 model summary for the Chile-CTPP Region FTA

| Dependent Variable:<br>Model: | Trade costs ($\tilde{t}_{i,j}$)<br>(3) |
|---|---|
| *Variables* | |
| Distance (log(DIST$_{ij}$)) | -0.5650*** |
| | (0.0290) |
| Contiguity (CNTG$_{ij}$) | 0.6239*** |
| | (0.1067) |
| Language (LANG$_{ij}$) | 0.3019*** |
| | (0.0471) |
| Colony (CLNY$_{ij}$) | 0.3234*** |
| | (0.0747) |
| *Fixed-effects* | |
| Exporter ($\pi_{i,t}$) | Yes |
| Importer ($\chi_{j,t}$) | Yes |
| *Fit statistics* | |
| Observations | 26,044 |
| Squared Correlation | 0.35098 |
| Pseudo R$^2$ | 0.19043 |
| BIC | 8,415.9 |

*Clustered standard-errors in parentheses*
*Signif. Codes:  ***: 0.01, **: 0.05, *: 0.1*

Column II of Table 9 reveals that, under the conditional general equilibrium scenario, the member countries of the Chile-CTPP Region FTA would experience the largest export increase, ranging from 2.69% (Malaysia) to 6.83% (Australia), of their total exports. Chile would increase its exports by 4.05%.

Column III of Table 9 reveals that the values of the Full Endowment general equilibrium effects of the FTAs above on exports are qualitatively identical to the corresponding conditional equilibrium effects, despite some quantitative differences.

Column IV of Table 9, the counterfactual analysis, suggests that the Full Endowment welfare effects of the Chile-CTPP Region FTA are favourable for its members, ranging from 0.82% (Malaysia) to 1.47% (Japan), and slightly negative or null for non-member countries. Chile would increase its GDP by 0.99%.

The decomposition of the Full Endowment GE effects reported in columns V, VI, and VII suggests that both consumers and producers in member countries of the FTAs face positive effects with lower inward multilateral resistances (for consumers) and lower outward multilateral resistances, which translate into higher factory-gate prices (for producers) relative to the effects on the consumers in the reference country (i.e., Germany for the current case).

# 5   Country tables

## 5.1   Conditional GE and Full Endownment GE simulation results for the Chile-USA FTA

Table 7: Conditional GE and Full Endownment GE simulation results for the Chile-USA FTA

| Exporter | Conditional GE | Full Endownment GE | | | | |
|---|---|---|---|---|---|---|
| | %Δ Trade | %Δ Trade | %Δ rGDP | %Δ IMR | %Δ OMR | %Δ Prices |
| ABW | 0.00 | 0.00 | 0.00 | 0.00 | 0.00 | 0.00 |
| AFG | 0.00 | 0.00 | 0.00 | 0.00 | 0.00 | 0.00 |
| AGO | 0.00 | 0.00 | 0.00 | 0.00 | 0.00 | 0.00 |
| AIA | 0.00 | -0.03 | -0.04 | 0.00 | 0.05 | -0.05 |
| ALB | 0.00 | 0.00 | 0.00 | 0.00 | 0.00 | 0.00 |
| AND | 0.00 | 0.00 | 0.00 | 0.00 | 0.00 | 0.00 |
| ANT | 0.00 | 0.00 | 0.00 | 0.00 | 0.00 | 0.00 |
| ARE | 0.00 | 0.00 | 0.00 | 0.00 | 0.00 | 0.00 |
| ARG | -0.05 | -0.06 | -0.01 | -0.01 | 0.02 | -0.01 |
| ARM | 0.00 | 0.00 | 0.00 | 0.00 | 0.00 | 0.00 |
| ATG | 0.00 | 0.00 | 0.00 | 0.00 | 0.00 | 0.00 |
| AUS | 0.00 | 0.00 | 0.00 | 0.00 | 0.00 | 0.00 |
| AUT | 0.00 | 0.00 | 0.00 | 0.00 | 0.00 | 0.00 |
| AZE | 0.00 | 0.00 | 0.00 | 0.00 | 0.00 | 0.00 |
| BDI | 0.00 | 0.00 | 0.00 | 0.00 | 0.00 | 0.00 |
| BEN | 0.00 | 0.00 | 0.00 | 0.00 | 0.00 | 0.00 |
| BFA | 0.00 | 0.00 | 0.00 | 0.00 | 0.00 | 0.00 |
| BGD | 0.00 | 0.00 | 0.00 | 0.00 | 0.00 | 0.00 |
| BGR | 0.00 | 0.00 | 0.00 | 0.00 | 0.00 | 0.00 |
| BHR | 0.00 | 0.00 | 0.00 | 0.00 | 0.00 | 0.00 |
| BHS | 0.00 | 0.00 | 0.00 | 0.00 | 0.00 | 0.00 |
| BIH | 0.00 | 0.00 | 0.00 | 0.00 | 0.00 | 0.00 |
| BLR | 0.00 | 0.00 | 0.00 | 0.00 | 0.00 | 0.00 |
| BLZ | 0.00 | 0.00 | 0.00 | 0.00 | 0.00 | 0.00 |
| BMU | 0.00 | 0.00 | 0.00 | 0.00 | 0.00 | 0.00 |
| BOL | -0.18 | -0.16 | -0.01 | 0.03 | -0.02 | 0.02 |
| BRA | -0.03 | -0.04 | 0.00 | 0.00 | 0.01 | 0.00 |
| BRB | 0.00 | 0.00 | 0.00 | 0.00 | 0.00 | 0.00 |
| BRN | 0.00 | 0.00 | 0.00 | 0.00 | 0.00 | 0.00 |
| BTN | 0.00 | 0.00 | 0.00 | 0.00 | 0.00 | 0.00 |
| BWA | 0.00 | 0.00 | 0.00 | 0.00 | 0.00 | 0.00 |
| CAF | 0.00 | 0.00 | 0.00 | 0.00 | 0.00 | 0.00 |
| CAN | -0.01 | -0.01 | 0.00 | 0.00 | 0.00 | 0.00 |
| CCK | 0.00 | 0.00 | 0.00 | 0.00 | 0.00 | 0.00 |
| CHE | 0.00 | 0.00 | 0.00 | 0.00 | 0.00 | 0.00 |
| **CHL** | **5.44** | **5.64** | **0.48** | **-0.29** | **-0.22** | **0.19** |
| CHN | 0.00 | -0.01 | 0.00 | 0.00 | 0.00 | 0.00 |
| CIV | 0.00 | 0.00 | 0.00 | 0.00 | 0.00 | 0.00 |
| CMR | 0.00 | 0.00 | 0.00 | 0.00 | 0.00 | 0.00 |
| COG | 0.00 | -0.01 | 0.00 | 0.00 | 0.01 | -0.01 |
| COK | 0.00 | 0.00 | 0.00 | 0.00 | 0.00 | 0.00 |
| COL | -0.04 | -0.03 | 0.00 | 0.00 | 0.00 | 0.00 |

| | | | | | |
|---|---|---|---|---|---|
| COM | 0.00 | 0.00 | 0.00 | 0.00 | 0.00 | 0.00 |
| CPV | 0.00 | 0.00 | 0.00 | 0.00 | 0.00 | 0.00 |
| CRI | -0.01 | -0.01 | 0.00 | 0.00 | 0.00 | 0.00 |
| CUB | 0.00 | 0.00 | 0.00 | 0.00 | 0.00 | 0.00 |
| CXR | 0.00 | 0.00 | 0.00 | 0.00 | 0.00 | 0.00 |
| CYM | 0.00 | 0.00 | 0.00 | 0.00 | 0.00 | 0.00 |
| CYP | 0.00 | 0.00 | 0.00 | 0.00 | 0.00 | 0.00 |
| CZE | 0.00 | 0.00 | 0.00 | 0.00 | 0.00 | 0.00 |
| DEU* | 0.00 | 0.00 | 0.00 | 0.00 | 0.00 | 0.00 |
| DJI | 0.00 | 0.00 | 0.00 | 0.00 | 0.00 | 0.00 |
| DMA | 0.00 | 0.00 | 0.00 | 0.00 | 0.00 | 0.00 |
| DNK | 0.00 | 0.00 | 0.00 | 0.00 | 0.00 | 0.00 |
| DOM | 0.00 | 0.00 | 0.00 | 0.00 | 0.00 | 0.00 |
| DZA | 0.00 | 0.00 | 0.00 | 0.00 | 0.00 | 0.00 |
| ECU | -0.07 | -0.06 | 0.00 | 0.01 | 0.00 | 0.00 |
| EGY | 0.00 | 0.00 | 0.00 | 0.00 | 0.00 | 0.00 |
| ERI | -0.01 | -0.01 | 0.00 | 0.00 | 0.00 | 0.00 |
| ESH | 0.00 | 0.00 | 0.00 | 0.00 | 0.00 | 0.00 |
| ESP | -0.01 | -0.01 | 0.00 | 0.00 | 0.00 | 0.00 |
| EST | 0.00 | 0.00 | 0.00 | 0.00 | 0.00 | 0.00 |
| ETH | 0.00 | 0.00 | 0.00 | 0.00 | 0.00 | 0.00 |
| FIN | 0.00 | 0.00 | 0.00 | 0.00 | 0.00 | 0.00 |
| FJI | 0.00 | 0.00 | 0.00 | 0.00 | 0.00 | 0.00 |
| FLK | 0.00 | 0.00 | 0.00 | 0.00 | 0.00 | 0.00 |
| FRA | 0.00 | 0.00 | 0.00 | 0.00 | 0.00 | 0.00 |
| FRO | 0.00 | 0.00 | 0.00 | 0.00 | 0.00 | 0.00 |
| FSM | 0.00 | 0.00 | 0.00 | 0.00 | 0.00 | 0.00 |
| GAB | 0.00 | 0.00 | 0.00 | 0.00 | 0.00 | 0.00 |
| GBR | 0.00 | 0.00 | 0.00 | 0.00 | 0.00 | 0.00 |
| GEO | 0.00 | 0.00 | 0.00 | 0.00 | 0.00 | 0.00 |
| GHA | 0.00 | 0.00 | 0.00 | 0.00 | 0.00 | 0.00 |
| GIB | 0.00 | 0.00 | 0.00 | 0.00 | 0.00 | 0.00 |
| GIN | 0.00 | 0.00 | 0.00 | 0.00 | 0.00 | 0.00 |
| GMB | 0.00 | 0.00 | 0.00 | 0.00 | 0.00 | 0.00 |
| GNB | 0.00 | 0.00 | 0.00 | 0.00 | 0.00 | 0.00 |
| GNQ | 0.00 | 0.00 | 0.00 | 0.00 | 0.00 | 0.00 |
| GRC | 0.00 | 0.00 | 0.00 | 0.00 | 0.00 | 0.00 |
| GRD | 0.00 | 0.00 | 0.00 | 0.00 | 0.00 | 0.00 |
| GRL | 0.00 | 0.00 | 0.00 | 0.00 | 0.00 | 0.00 |
| GTM | -0.01 | -0.01 | 0.00 | 0.00 | 0.00 | 0.00 |
| GUY | 0.00 | 0.00 | 0.00 | 0.00 | 0.00 | 0.00 |
| HKG | 0.00 | 0.00 | 0.00 | 0.00 | 0.00 | 0.00 |
| HND | -0.01 | -0.01 | 0.00 | 0.00 | 0.00 | 0.00 |
| HRV | 0.00 | 0.00 | 0.00 | 0.00 | 0.00 | 0.00 |
| HTI | -0.01 | -0.01 | 0.00 | 0.00 | 0.00 | 0.00 |
| HUN | 0.00 | 0.00 | 0.00 | 0.00 | 0.00 | 0.00 |
| IDN | 0.00 | 0.00 | 0.00 | 0.00 | 0.00 | 0.00 |
| IND | 0.00 | 0.00 | 0.00 | 0.00 | 0.00 | 0.00 |
| IRL | 0.00 | 0.00 | 0.00 | 0.00 | 0.00 | 0.00 |
| IRN | 0.00 | 0.00 | 0.00 | 0.00 | 0.00 | 0.00 |
| IRQ | 0.00 | 0.00 | 0.00 | 0.00 | 0.00 | 0.00 |
| ISL | -0.01 | -0.01 | 0.00 | 0.00 | 0.00 | 0.00 |
| ISR | 0.00 | 0.00 | 0.00 | 0.00 | 0.00 | 0.00 |
| ITA | 0.00 | 0.00 | 0.00 | 0.00 | 0.00 | 0.00 |
| JAM | 0.00 | 0.00 | 0.00 | 0.00 | 0.00 | 0.00 |
| JOR | 0.00 | 0.00 | 0.00 | 0.00 | 0.00 | 0.00 |
| JPN | 0.00 | 0.00 | 0.00 | 0.00 | 0.00 | 0.00 |
| KAZ | 0.00 | 0.00 | 0.00 | 0.00 | 0.00 | 0.00 |
| KEN | 0.00 | 0.00 | 0.00 | 0.00 | 0.00 | 0.00 |
| KGZ | 0.00 | 0.00 | 0.00 | 0.00 | 0.00 | 0.00 |
| KHM | 0.00 | 0.00 | 0.00 | 0.00 | 0.00 | 0.00 |
| KIR | 0.00 | 0.00 | 0.00 | 0.00 | 0.00 | 0.00 |
| KNA | 0.00 | 0.00 | 0.00 | 0.00 | 0.00 | 0.00 |
| KOR | -0.01 | -0.01 | 0.00 | 0.00 | 0.00 | 0.00 |
| KWT | 0.00 | 0.00 | 0.00 | 0.00 | 0.00 | 0.00 |
| LAO | 0.00 | 0.00 | 0.00 | 0.00 | 0.00 | 0.00 |
| LBN | 0.00 | 0.00 | 0.00 | 0.00 | 0.00 | 0.00 |
| LBR | 0.00 | 0.00 | 0.00 | 0.00 | 0.00 | 0.00 |
| LBY | 0.00 | 0.00 | 0.00 | 0.00 | 0.00 | 0.00 |
| LCA | 0.00 | 0.00 | 0.00 | 0.00 | 0.00 | 0.00 |
| LKA | 0.00 | 0.00 | 0.00 | 0.00 | 0.00 | 0.00 |
| LSO | 0.00 | 0.00 | 0.00 | 0.00 | 0.00 | 0.00 |
| LTU | 0.00 | 0.00 | 0.00 | 0.00 | 0.00 | 0.00 |
| LVA | 0.00 | 0.00 | 0.00 | 0.00 | 0.00 | 0.00 |
| MAC | 0.00 | 0.00 | 0.00 | 0.00 | 0.00 | 0.00 |
| MAR | 0.00 | 0.00 | 0.00 | 0.00 | 0.00 | 0.00 |
| MDA | 0.00 | 0.00 | 0.00 | 0.00 | 0.00 | 0.00 |
| MDG | 0.00 | 0.00 | 0.00 | 0.00 | 0.00 | 0.00 |
| MDV | 0.00 | 0.00 | 0.00 | 0.00 | 0.00 | 0.00 |
| MEX | -0.01 | -0.01 | 0.00 | 0.00 | 0.00 | 0.00 |
| MHL | 0.00 | 0.00 | 0.00 | 0.00 | 0.00 | 0.00 |
| MKD | 0.00 | 0.00 | 0.00 | 0.00 | 0.00 | 0.00 |
| MLI | 0.00 | 0.00 | 0.00 | 0.00 | 0.00 | 0.00 |
| MLT | 0.00 | 0.00 | 0.00 | 0.00 | 0.00 | 0.00 |
| MMR | 0.00 | 0.00 | 0.00 | 0.00 | 0.00 | 0.00 |
| MNG | 0.00 | 0.00 | 0.00 | 0.00 | 0.00 | 0.00 |
| MNP | 0.00 | 0.00 | 0.00 | 0.00 | 0.00 | 0.00 |
| MOZ | 0.00 | 0.00 | 0.00 | 0.00 | 0.00 | 0.00 |
| MRT | 0.00 | 0.00 | 0.00 | 0.00 | 0.00 | 0.00 |
| MSR | 0.00 | 0.00 | 0.00 | 0.00 | 0.00 | 0.00 |
| MUS | 0.00 | 0.00 | 0.00 | 0.00 | 0.00 | 0.00 |
| MWI | 0.00 | 0.00 | 0.00 | 0.00 | 0.00 | 0.00 |
| MYS | 0.00 | 0.00 | 0.00 | 0.00 | 0.00 | 0.00 |
| NAM | 0.00 | 0.00 | 0.00 | 0.00 | 0.00 | 0.00 |
| NCL | 0.00  | 0.00  | 0.00  | 0.00 | 0.00 | 0.00  |
| NER | 0.00  | 0.00  | 0.00  | 0.00 | 0.00 | 0.00  |
| NFK | 0.00  | 0.00  | 0.00  | 0.00 | 0.00 | 0.00  |
| NGA | -0.01 | -0.01 | 0.00  | 0.00 | 0.00 | 0.00  |
| NIC | 0.00  | 0.00  | 0.00  | 0.00 | 0.00 | 0.00  |
| NIU | 0.00  | 0.00  | 0.00  | 0.00 | 0.00 | 0.00  |
| NLD | 0.00  | 0.00  | 0.00  | 0.00 | 0.00 | 0.00  |
| NOR | 0.00  | 0.00  | 0.00  | 0.00 | 0.00 | 0.00  |
| NPL | 0.00  | 0.00  | 0.00  | 0.00 | 0.00 | 0.00  |
| NRU | 0.00  | 0.00  | 0.00  | 0.00 | 0.00 | 0.00  |
| NZL | 0.00  | 0.00  | 0.00  | 0.00 | 0.00 | 0.00  |
| OMN | 0.00  | 0.00  | 0.00  | 0.00 | 0.00 | 0.00  |
| PAK | 0.00  | 0.00  | 0.00  | 0.00 | 0.00 | 0.00  |
| PAN | 0.00  | 0.00  | 0.00  | 0.00 | 0.00 | 0.00  |
| PCN | 0.00  | 0.00  | 0.00  | 0.00 | 0.00 | 0.00  |
| PER | -0.06 | -0.06 | 0.00  | 0.01 | 0.00 | 0.00  |
| PHL | 0.00  | 0.00  | 0.00  | 0.00 | 0.00 | 0.00  |
| PLW | 0.00  | 0.00  | 0.00  | 0.00 | 0.00 | 0.00  |
| PNG | 0.00  | 0.00  | 0.00  | 0.00 | 0.00 | 0.00  |
| POL | 0.00  | 0.00  | 0.00  | 0.00 | 0.00 | 0.00  |
| PRK | 0.00  | 0.00  | 0.00  | 0.00 | 0.00 | 0.00  |
| PRT | 0.00  | 0.00  | 0.00  | 0.00 | 0.00 | 0.00  |
| PRY | 0.00  | -0.02 | -0.02 | 0.00 | 0.02 | -0.02 |
| PYF | 0.00  | 0.00  | 0.00  | 0.00 | 0.00 | 0.00  |
| QAT | 0.00  | 0.00  | 0.00  | 0.00 | 0.00 | 0.00  |
| ROM | 0.00  | 0.00  | 0.00  | 0.00 | 0.00 | 0.00  |
| RUS | 0.00  | 0.00  | 0.00  | 0.00 | 0.00 | 0.00  |
| RWA | 0.00  | 0.00  | 0.00  | 0.00 | 0.00 | 0.00  |
| SAU | 0.00  | 0.00  | 0.00  | 0.00 | 0.00 | 0.00  |
| SDN | 0.00  | 0.00  | 0.00  | 0.00 | 0.00 | 0.00  |
| SEN | 0.00  | 0.00  | 0.00  | 0.00 | 0.00 | 0.00  |
| SGP | 0.00  | 0.00  | 0.00  | 0.00 | 0.00 | 0.00  |
| SHN | 0.00  | 0.00  | -0.01 | 0.01 | 0.00 | 0.00  |
| SLB | 0.00  | 0.00  | 0.00  | 0.00 | 0.00 | 0.00  |
| SLE | 0.00  | 0.00  | 0.00  | 0.00 | 0.00 | 0.00  |
| SLV | 0.00  | 0.00  | 0.00  | 0.00 | 0.00 | 0.00  |
| SMR | 0.00  | 0.00  | 0.00  | 0.00 | 0.00 | 0.00  |
| SOM | 0.00  | 0.00  | -0.01 | 0.01 | 0.00 | 0.00  |
| SPM | 0.00  | 0.00  | 0.00  | 0.00 | 0.00 | 0.00  |
| STP | 0.00  | 0.00  | 0.00  | 0.00 | 0.00 | 0.00  |
| SUR | 0.00  | 0.00  | 0.00  | 0.00 | 0.00 | 0.00  |
| SVK | 0.00  | 0.00  | 0.00  | 0.00 | 0.00 | 0.00  |
| SVN | 0.00  | 0.00  | 0.00  | 0.00 | 0.00 | 0.00  |
| SWE | 0.00  | 0.00  | 0.00  | 0.00 | 0.00 | 0.00  |
| SWZ | 0.00  | 0.00  | 0.00  | 0.00 | 0.00 | 0.00  |
| SYC | 0.00  | 0.00  | 0.00  | 0.00 | 0.00 | 0.00  |
| SYR | 0.00  | 0.00  | 0.00  | 0.00 | 0.00 | 0.00  |

| | | | | | |
|---|---|---|---|---|---|
| TCA | 0.00 | 0.00 | 0.00 | 0.00 | 0.00 | 0.00 |
| TCD | 0.00 | 0.00 | -0.01 | 0.01 | 0.00 | 0.00 |
| TGO | 0.00 | 0.00 | 0.00 | 0.00 | 0.00 | 0.00 |
| THA | 0.00 | 0.00 | 0.00 | 0.00 | 0.00 | 0.00 |
| TJK | 0.00 | 0.00 | 0.00 | 0.00 | 0.00 | 0.00 |
| TKL | 0.00 | 0.00 | 0.00 | 0.00 | 0.00 | 0.00 |
| TKM | 0.00 | 0.00 | 0.00 | 0.00 | 0.00 | 0.00 |
| TMP | 0.00 | 0.00 | 0.00 | 0.00 | 0.00 | 0.00 |
| TON | -0.01 | -0.01 | 0.00 | 0.00 | 0.00 | 0.00 |
| TTO | 0.00 | 0.00 | 0.00 | 0.00 | 0.00 | 0.00 |
| TUN | 0.00 | 0.00 | 0.00 | 0.00 | 0.00 | 0.00 |
| TUR | 0.00 | 0.00 | 0.00 | 0.00 | 0.00 | 0.00 |
| TUV | 0.00 | 0.00 | -0.01 | 0.01 | 0.00 | 0.00 |
| TWN | 0.00 | 0.00 | 0.00 | 0.00 | 0.00 | 0.00 |
| TZA | 0.00 | 0.00 | 0.00 | 0.00 | 0.00 | 0.00 |
| UGA | 0.00 | 0.00 | 0.00 | 0.00 | 0.00 | 0.00 |
| UKR | 0.00 | 0.00 | 0.00 | 0.00 | 0.00 | 0.00 |
| URY | -0.03 | -0.03 | 0.00 | 0.00 | 0.01 | 0.00 |
| **USA** | **0.23** | **0.23** | **0.01** | **0.00** | **0.00** | **0.00** |
| UZB | 0.00 | 0.00 | 0.00 | 0.00 | 0.00 | 0.00 |
| VCT | 0.00 | 0.00 | 0.00 | 0.00 | 0.00 | 0.00 |
| VEN | -0.01 | -0.01 | 0.00 | 0.00 | 0.00 | 0.00 |
| VGB | 0.00 | 0.00 | 0.00 | 0.00 | 0.00 | 0.00 |
| VNM | 0.00 | 0.00 | 0.00 | 0.00 | 0.00 | 0.00 |
| VUT | 0.00 | 0.00 | 0.00 | 0.00 | 0.00 | 0.00 |
| WLF | 0.00 | 0.00 | 0.00 | 0.00 | 0.00 | 0.00 |
| WSM | 0.00 | 0.00 | 0.00 | 0.00 | 0.00 | 0.00 |
| YEM | 0.00 | 0.00 | 0.00 | 0.00 | 0.00 | 0.00 |
| ZAF | 0.00 | 0.00 | 0.00 | 0.00 | 0.00 | 0.00 |
| ZAR | 0.00 | 0.00 | 0.00 | 0.00 | 0.00 | 0.00 |
| ZMB | 0.00 | 0.00 | 0.00 | 0.00 | 0.00 | 0.00 |
| ZWE | 0.00 | 0.00 | 0.00 | 0.00 | 0.00 | 0.00 |

## 5.2 Conditional GE and Full Endownment GE simulation results for the Chile-China FTA

Table 8: Conditional GE and Full Endownment GE simulation results for the Chile-China FTA

| Exporter | Conditional GE | Full Endownment GE | | | | |
|---|---|---|---|---|---|---|
| | %Δ Trade | %Δ Trade | %Δ rGDP | %Δ IMR | %Δ OMR | %Δ Prices |
| ABW | 0.00 | 0.00 | 0.00 | 0.00 | 0.00 | 0.00 |
| AFG | 0.00 | 0.00 | 0.00 | 0.00 | 0.00 | 0.00 |
| AGO | 0.00 | 0.00 | 0.00 | 0.00 | 0.00 | 0.00 |
| AIA | 0.00 | 0.00 | 0.00 | 0.00 | 0.00 | 0.00 |
| ALB | 0.00 | 0.00 | 0.00 | 0.00 | 0.00 | 0.00 |
| AND | 0.00 | 0.00 | 0.00 | 0.00 | 0.00 | 0.00 |
| ANT | 0.00 | 0.00 | 0.00 | 0.00 | 0.00 | 0.00 |
| ARE | 0.00 | 0.00 | 0.00 | 0.00 | 0.00 | 0.00 |
| ARG | -0.01 | -0.01 | 0.00 | 0.00 | 0.00 | 0.00 |
| ARM | 0.00 | 0.00 | 0.00 | 0.00 | 0.00 | 0.00 |
| ATG | 0.00 | 0.00 | 0.00 | 0.00 | 0.00 | 0.00 |
| AUS | 0.00 | 0.00 | 0.00 | 0.00 | 0.00 | 0.00 |
| AUT | 0.00 | 0.00 | 0.00 | 0.00 | 0.00 | 0.00 |
| AZE | 0.00 | 0.00 | 0.00 | 0.00 | 0.00 | 0.00 |
| BDI | 0.00 | 0.00 | 0.00 | 0.00 | 0.00 | 0.00 |
| BEN | 0.00 | 0.00 | 0.00 | 0.00 | 0.00 | 0.00 |
| BFA | 0.00 | 0.00 | 0.00 | 0.00 | 0.00 | 0.00 |
| BGD | 0.00 | 0.00 | 0.00 | 0.00 | 0.00 | 0.00 |
| BGR | 0.00 | 0.00 | 0.00 | 0.00 | 0.00 | 0.00 |
| BHR | 0.00 | 0.00 | 0.00 | 0.00 | 0.00 | 0.00 |
| BHS | 0.00 | 0.00 | 0.00 | 0.00 | 0.00 | 0.00 |
| BIH | 0.00 | 0.00 | 0.00 | 0.00 | 0.00 | 0.00 |
| BLZ | 0.00 | 0.00 | 0.00 | 0.00 | 0.00 | 0.00 |
| BMU | 0.00 | 0.00 | 0.00 | 0.00 | 0.00 | 0.00 |
| BOL | -0.14 | -0.12 | -0.01 | 0.03 | -0.02 | 0.02 |
| BRA | -0.01 | -0.01 | 0.00 | 0.00 | 0.00 | 0.00 |
| BRB | 0.00 | 0.00 | 0.00 | 0.00 | 0.00 | 0.00 |
| BRN | 0.00 | 0.00 | 0.00 | 0.00 | 0.00 | 0.00 |
| BTN | 0.00 | 0.00 | 0.00 | 0.00 | 0.00 | 0.00 |
| BWA | 0.00 | 0.00 | 0.00 | 0.00 | 0.00 | 0.00 |
| CAF | 0.00 | 0.00 | 0.00 | 0.00 | 0.00 | 0.00 |
| CAN | 0.00 | 0.00 | 0.00 | 0.00 | 0.00 | 0.00 |
| CCK | 0.00 | 0.00 | 0.00 | 0.00 | 0.00 | 0.00 |
| CHE | 0.00 | 0.00 | 0.00 | 0.00 | 0.00 | 0.00 |
| **CHL** | **2.01** | **2.15** | **0.22** | **-0.08** | **-0.17** | **0.14** |
| **CHN** | **0.09** | **0.08** | **0.00** | **-0.01** | **0.00** | **0.00** |
| CIV | 0.00 | 0.00 | 0.00 | 0.00 | 0.00 | 0.00 |
| CMR | 0.00 | 0.00 | 0.00 | 0.00 | 0.00 | 0.00 |
| COG | 0.00 | 0.00 | 0.00 | 0.00 | 0.00 | 0.00 |
| COK | 0.00 | 0.00 | 0.00 | 0.00 | 0.00 | 0.00 |
| COL | -0.01 | -0.01 | 0.00 | 0.00 | 0.00 | 0.00 |
| COM | 0.00 | 0.00 | 0.00 | 0.00 | 0.00 | 0.00 |
| CPV | 0.00 | 0.00 | 0.00 | 0.00 | 0.00 | 0.00 |
| CRI | -0.01 | -0.01 | 0.00 | 0.00 | 0.00 | 0.00 |
| CUB | 0.00 | 0.00 | 0.00 | 0.00 | 0.00 | 0.00 |
| CXR | 0.00 | 0.00 | 0.00 | 0.00 | 0.00 | 0.00 |
| CYM | 0.00 | 0.00 | 0.00 | 0.00 | 0.00 | 0.00 |
| CYP | 0.00 | 0.00 | 0.00 | 0.00 | 0.00 | 0.00 |
| CZE | 0.00 | 0.00 | 0.00 | 0.00 | 0.00 | 0.00 |
| DEU* | 0.00 | 0.00 | 0.00 | 0.00 | 0.00 | 0.00 |
| DJI | 0.00 | 0.00 | 0.00 | 0.00 | 0.00 | 0.00 |

| | | | | | |
|---|---|---|---|---|---|
| DMA | 0.00 | 0.00 | 0.00 | 0.00 | 0.00 | 0.00 |
| DNK | 0.00 | 0.00 | 0.00 | 0.00 | 0.00 | 0.00 |
| DOM | 0.00 | 0.00 | 0.00 | 0.00 | 0.00 | 0.00 |
| DZA | 0.00 | 0.00 | 0.00 | 0.00 | 0.00 | 0.00 |
| ECU | -0.03 | -0.02 | 0.00 | 0.00 | 0.00 | 0.00 |
| EGY | 0.00 | 0.00 | 0.00 | 0.00 | 0.00 | 0.00 |
| ERI | 0.00 | 0.00 | 0.00 | 0.00 | 0.00 | 0.00 |
| ESH | 0.00 | 0.00 | 0.00 | 0.00 | 0.00 | 0.00 |
| ESP | 0.00 | 0.00 | 0.00 | 0.00 | 0.00 | 0.00 |
| EST | 0.00 | 0.00 | 0.00 | 0.00 | 0.00 | 0.00 |
| ETH | 0.00 | 0.00 | 0.00 | 0.00 | 0.00 | 0.00 |
| FIN | 0.00 | 0.00 | 0.00 | 0.00 | 0.00 | 0.00 |
| FJI | 0.00 | 0.00 | 0.00 | 0.00 | 0.00 | 0.00 |
| FLK | 0.00 | 0.00 | 0.00 | 0.00 | 0.00 | 0.00 |
| FRA | 0.00 | 0.00 | 0.00 | 0.00 | 0.00 | 0.00 |
| FRO | 0.00 | 0.00 | 0.00 | 0.00 | 0.00 | 0.00 |
| FSM | 0.00 | 0.00 | 0.00 | 0.00 | 0.00 | 0.00 |
| GAB | 0.00 | 0.00 | 0.00 | 0.00 | 0.00 | 0.00 |
| GBR | 0.00 | 0.00 | 0.00 | 0.00 | 0.00 | 0.00 |
| GEO | 0.00 | 0.00 | 0.00 | 0.00 | 0.00 | 0.00 |
| GHA | 0.00 | 0.00 | 0.00 | 0.00 | 0.00 | 0.00 |
| GIB | 0.00 | 0.00 | 0.00 | 0.00 | 0.00 | 0.00 |
| GIN | 0.00 | 0.00 | 0.00 | 0.00 | 0.00 | 0.00 |
| GMB | 0.00 | 0.00 | 0.00 | 0.00 | 0.00 | 0.00 |
| GNB | 0.00 | 0.00 | 0.00 | 0.00 | 0.00 | 0.00 |
| GNQ | 0.00 | 0.00 | 0.00 | 0.00 | 0.00 | 0.00 |
| GRC | 0.00 | 0.00 | 0.00 | 0.00 | 0.00 | 0.00 |
| GRD | 0.00 | 0.00 | 0.00 | 0.00 | 0.00 | 0.00 |
| GRL | 0.00 | 0.00 | 0.00 | 0.00 | 0.00 | 0.00 |
| GTM | -0.01 | 0.00 | 0.00 | 0.00 | 0.00 | 0.00 |
| GUY | 0.00 | 0.00 | 0.00 | 0.00 | 0.00 | 0.00 |
| HKG | 0.00 | 0.00 | 0.00 | 0.00 | 0.00 | 0.00 |
| HND | 0.00 | 0.00 | 0.00 | 0.00 | 0.00 | 0.00 |
| HRV | 0.00 | 0.00 | 0.00 | 0.00 | 0.00 | 0.00 |
| HTI | 0.00 | 0.00 | 0.00 | 0.00 | 0.00 | 0.00 |
| HUN | 0.00 | 0.00 | 0.00 | 0.00 | 0.00 | 0.00 |
| IDN | 0.00 | 0.00 | 0.00 | 0.00 | 0.00 | 0.00 |
| IND | 0.00 | 0.00 | 0.00 | 0.00 | 0.00 | 0.00 |
| IRL | 0.00 | 0.00 | 0.00 | 0.00 | 0.00 | 0.00 |
| IRN | 0.00 | 0.00 | 0.00 | 0.00 | 0.00 | 0.00 |
| IRQ | 0.00 | -0.01 | 0.00 | 0.00 | 0.00 | 0.00 |
| ISL | 0.00 | 0.00 | 0.00 | 0.00 | 0.00 | 0.00 |
| ISR | 0.00 | 0.00 | 0.00 | 0.00 | 0.00 | 0.00 |
| ITA | 0.00 | 0.00 | 0.00 | 0.00 | 0.00 | 0.00 |
| JAM | 0.00 | 0.00 | 0.00 | 0.00 | 0.00 | 0.00 |
| JOR | 0.00 | 0.00 | 0.00 | 0.00 | 0.00 | 0.00 |
| JPN | 0.00 | 0.00 | 0.00 | 0.00 | 0.00 | 0.00 |

| | | | | | |
|---|---|---|---|---|---|
| KAZ | 0.00 | 0.00 | 0.00 | 0.00 | 0.00 | 0.00 |
| KEN | 0.00 | 0.00 | 0.00 | 0.00 | 0.00 | 0.00 |
| KGZ | 0.00 | 0.00 | 0.00 | 0.00 | 0.00 | 0.00 |
| KHM | 0.00 | 0.00 | 0.00 | 0.00 | 0.00 | 0.00 |
| KIR | 0.00 | 0.00 | 0.00 | 0.00 | 0.00 | 0.00 |
| KNA | 0.00 | 0.00 | 0.00 | 0.00 | 0.00 | 0.00 |
| KOR | -0.01 | -0.01 | 0.00 | 0.00 | 0.00 | 0.00 |
| KWT | 0.00 | 0.00 | 0.00 | 0.00 | 0.00 | 0.00 |
| LAO | 0.00 | 0.00 | 0.00 | 0.00 | 0.00 | 0.00 |
| LBN | 0.00 | 0.00 | 0.00 | 0.00 | 0.00 | 0.00 |
| LBR | 0.00 | 0.00 | 0.00 | 0.00 | 0.00 | 0.00 |
| LBY | 0.00 | 0.00 | 0.00 | 0.00 | 0.00 | 0.00 |
| LCA | 0.00 | 0.00 | 0.00 | 0.00 | 0.00 | 0.00 |
| LKA | 0.00 | 0.00 | 0.00 | 0.00 | 0.00 | 0.00 |
| LSO | 0.00 | 0.00 | 0.00 | 0.00 | 0.00 | 0.00 |
| LTU | 0.00 | 0.00 | 0.00 | 0.00 | 0.00 | 0.00 |
| LVA | 0.00 | 0.00 | 0.00 | 0.00 | 0.00 | 0.00 |
| MAC | 0.00 | 0.00 | 0.00 | 0.00 | 0.00 | 0.00 |
| MAR | 0.00 | 0.00 | 0.00 | 0.00 | 0.00 | 0.00 |
| MDA | 0.00 | 0.00 | 0.00 | 0.00 | 0.00 | 0.00 |
| MDG | 0.00 | 0.00 | 0.00 | 0.00 | 0.00 | 0.00 |
| MDV | 0.00 | 0.00 | 0.00 | 0.00 | 0.00 | 0.00 |
| MEX | 0.00 | 0.00 | 0.00 | 0.00 | 0.00 | 0.00 |
| MHL | 0.00 | 0.00 | 0.00 | 0.00 | 0.00 | 0.00 |
| MKD | 0.00 | 0.00 | 0.00 | 0.00 | 0.00 | 0.00 |
| MLI | 0.00 | 0.00 | 0.00 | 0.00 | 0.00 | 0.00 |
| MLT | 0.00 | 0.00 | 0.00 | 0.00 | 0.00 | 0.00 |
| MMR | 0.00 | 0.00 | 0.00 | 0.00 | 0.00 | 0.00 |
| MNG | 0.00 | 0.00 | 0.00 | 0.00 | 0.00 | 0.00 |
| MNP | 0.00 | 0.00 | 0.00 | 0.00 | 0.00 | 0.00 |
| MOZ | 0.00 | 0.00 | 0.00 | 0.00 | 0.00 | 0.00 |
| MRT | 0.00 | 0.00 | 0.00 | 0.00 | 0.00 | 0.00 |
| MSR | 0.00 | 0.00 | 0.00 | 0.00 | 0.00 | 0.00 |
| MUS | 0.00 | 0.00 | 0.00 | 0.00 | 0.00 | 0.00 |
| MWI | 0.00 | 0.00 | 0.00 | 0.00 | 0.00 | 0.00 |
| MYS | 0.00 | 0.00 | 0.00 | 0.00 | 0.00 | 0.00 |
| NAM | 0.00 | 0.00 | 0.00 | 0.00 | 0.00 | 0.00 |
| NCL | 0.00 | 0.00 | 0.00 | 0.00 | 0.00 | 0.00 |
| NER | 0.00 | 0.00 | 0.00 | 0.00 | 0.00 | 0.00 |
| NFK | 0.00 | 0.00 | 0.00 | 0.00 | 0.00 | 0.00 |
| NGA | 0.00 | 0.00 | 0.00 | 0.00 | 0.00 | 0.00 |
| NIC | 0.00 | 0.00 | 0.00 | 0.00 | 0.00 | 0.00 |
| NIU | 0.00 | 0.00 | 0.00 | 0.00 | 0.00 | 0.00 |
| NLD | 0.00 | 0.00 | 0.00 | 0.00 | 0.00 | 0.00 |
| NOR | 0.00 | 0.00 | 0.00 | 0.00 | 0.00 | 0.00 |
| NPL | 0.00 | 0.00 | 0.00 | 0.00 | 0.00 | 0.00 |
| NRU | 0.00 | 0.00 | 0.00 | 0.00 | 0.00 | 0.00 |
| NZL | 0.00 | 0.00 | 0.00 | 0.00 | 0.00 | 0.00 |
| OMN | 0.00 | 0.00 | 0.00 | 0.00 | 0.00 | 0.00 |
| PAK | 0.00 | 0.00 | 0.00 | 0.00 | 0.00 | 0.00 |
| PAN | 0.00 | 0.00 | 0.00 | 0.00 | 0.00 | 0.00 |
| PCN | 0.00 | 0.00 | 0.00 | 0.00 | 0.00 | 0.00 |
| PER | -0.04 | -0.03 | 0.00 | 0.01 | -0.01 | 0.00 |
| PHL | 0.00 | 0.00 | 0.00 | 0.00 | 0.00 | 0.00 |
| PLW | 0.00 | 0.00 | 0.00 | 0.00 | 0.00 | 0.00 |
| PNG | 0.00 | 0.00 | 0.00 | 0.00 | 0.00 | 0.00 |
| POL | 0.00 | 0.00 | 0.00 | 0.00 | 0.00 | 0.00 |
| PRK | 0.00 | 0.00 | 0.00 | 0.00 | 0.00 | 0.00 |
| PRT | 0.00 | 0.00 | 0.00 | 0.00 | 0.00 | 0.00 |
| PRY | -0.02 | -0.01 | -0.01 | 0.00 | 0.00 | 0.00 |
| PYF | 0.00 | 0.00 | 0.00 | 0.00 | 0.00 | 0.00 |
| QAT | 0.00 | 0.00 | 0.00 | 0.00 | 0.00 | 0.00 |
| ROM | 0.00 | 0.00 | 0.00 | 0.00 | 0.00 | 0.00 |
| RUS | 0.00 | 0.00 | 0.00 | 0.00 | 0.00 | 0.00 |
| RWA | 0.00 | 0.00 | 0.00 | 0.00 | 0.00 | 0.00 |
| SAU | 0.00 | 0.00 | 0.00 | 0.00 | 0.00 | 0.00 |
| SDN | 0.00 | 0.00 | 0.00 | 0.00 | 0.00 | 0.00 |
| SEN | 0.00 | 0.00 | 0.00 | 0.00 | 0.00 | 0.00 |
| SGP | 0.00 | 0.00 | 0.00 | 0.00 | 0.00 | 0.00 |
| SHN | 0.00 | 0.00 | 0.00 | 0.00 | 0.00 | 0.00 |
| SLB | 0.00 | 0.00 | 0.00 | 0.00 | 0.00 | 0.00 |
| SLE | 0.00 | 0.00 | 0.00 | 0.00 | 0.00 | 0.00 |
| SLV | 0.00 | 0.00 | 0.00 | 0.00 | 0.00 | 0.00 |
| SMR | 0.00 | 0.00 | 0.00 | 0.00 | 0.00 | 0.00 |
| SOM | 0.00 | 0.00 | 0.00 | 0.00 | 0.00 | 0.00 |
| SPM | 0.00 | 0.00 | 0.00 | 0.00 | 0.00 | 0.00 |
| STP | 0.00 | 0.00 | 0.00 | 0.00 | 0.00 | 0.00 |
| SUR | 0.00 | 0.00 | 0.00 | 0.00 | 0.00 | 0.00 |
| SVK | 0.00 | 0.00 | 0.00 | 0.00 | 0.00 | 0.00 |
| SVN | 0.00 | 0.00 | 0.00 | 0.00 | 0.00 | 0.00 |
| SWE | 0.00 | 0.00 | 0.00 | 0.00 | 0.00 | 0.00 |
| SWZ | 0.00 | 0.00 | 0.00 | 0.00 | 0.00 | 0.00 |
| SYC | 0.00 | 0.00 | 0.00 | 0.00 | 0.00 | 0.00 |
| SYR | 0.00 | 0.00 | 0.00 | 0.00 | 0.00 | 0.00 |
| TCA | 0.00 | 0.00 | 0.00 | 0.00 | 0.00 | 0.00 |
| TCD | 0.00 | 0.00 | 0.00 | 0.00 | 0.00 | 0.00 |
| TGO | 0.00 | 0.00 | 0.00 | 0.00 | 0.00 | 0.00 |
| THA | 0.00 | 0.00 | 0.00 | 0.00 | 0.00 | 0.00 |
| TJK | 0.00 | 0.00 | 0.00 | 0.00 | 0.00 | 0.00 |
| TKL | 0.00 | 0.00 | 0.00 | 0.00 | 0.00 | 0.00 |
| TKM | 0.00 | 0.00 | 0.00 | 0.00 | 0.00 | 0.00 |
| TMP | 0.00 | 0.00 | 0.00 | 0.00 | 0.00 | 0.00 |
| TON | 0.00 | 0.00 | 0.00 | 0.00 | 0.00 | 0.00 |
| TTO | 0.00 | 0.00 | 0.00 | 0.00 | 0.00 | 0.00 |

| | | | | | |
|---|---|---|---|---|---|
| TUN | 0.00 | 0.00 | 0.00 | 0.00 | 0.00 | 0.00 |
| TUR | 0.00 | 0.00 | 0.00 | 0.00 | 0.00 | 0.00 |
| TUV | 0.00 | 0.05 | -0.02 | 0.01 | 0.01 | -0.01 |
| TZA | 0.00 | 0.00 | 0.00 | 0.00 | 0.00 | 0.00 |
| UGA | 0.00 | 0.00 | 0.00 | 0.00 | 0.00 | 0.00 |
| UKR | 0.00 | 0.00 | 0.00 | 0.00 | 0.00 | 0.00 |
| URY | -0.01 | -0.01 | 0.00 | 0.00 | 0.00 | 0.00 |
| USA | 0.00 | 0.00 | 0.00 | 0.00 | 0.00 | 0.00 |
| UZB | 0.00 | 0.00 | 0.00 | 0.00 | 0.00 | 0.00 |
| VCT | 0.00 | 0.00 | 0.00 | 0.00 | 0.00 | 0.00 |
| VEN | -0.01 | -0.01 | 0.00 | 0.00 | 0.00 | 0.00 |
| VGB | 0.00 | 0.00 | 0.00 | 0.00 | 0.00 | 0.00 |
| VNM | 0.00 | 0.00 | 0.00 | 0.00 | 0.00 | 0.00 |
| VUT | 0.00 | 0.00 | 0.00 | 0.00 | 0.00 | 0.00 |
| WLF | 0.00 | 0.00 | 0.00 | 0.00 | 0.00 | 0.00 |
| WSM | 0.00 | 0.00 | 0.00 | 0.00 | 0.00 | 0.00 |
| YEM | 0.00 | 0.00 | 0.00 | 0.00 | 0.00 | 0.00 |
| ZAF | 0.00 | 0.00 | 0.00 | 0.00 | 0.00 | 0.00 |
| ZAR | 0.00 | 0.00 | 0.00 | 0.00 | 0.00 | 0.00 |
| ZMB | 0.00 | 0.00 | 0.00 | 0.00 | 0.00 | 0.00 |
| ZWE | 0.00 | 0.00 | 0.00 | 0.00 | 0.00 | 0.00 |

## 5.3 Conditional GE and Full Endownment GE simulation results for the Chile-CTPP Region FTA

Table 9: Conditional GE and Full Endownment GE simulation results for the Chile-CTPP Region FTA

| Exporter | Conditional GE | Full Endownment GE | | | | |
|---|---|---|---|---|---|---|
| | %Δ Trade | %Δ Trade | %Δ rGDP | %Δ IMR | %Δ OMR | %Δ Prices |
| ABW | 0.00 | -0.04 | -0.02 | 0.06 | -0.04 | 0.03 |
| AFG | -0.01 | -0.03 | 0.00 | 0.01 | -0.01 | 0.01 |
| AGO | -0.03 | -0.06 | 0.00 | 0.01 | -0.01 | 0.01 |
| AIA | 0.00 | 0.00 | -0.04 | 0.04 | 0.01 | -0.01 |
| ALB | 0.01 | -0.01 | 0.00 | 0.00 | 0.01 | -0.01 |
| AND | 0.00 | -0.01 | 0.00 | -0.01 | 0.00 | 0.00 |
| ARE | 0.10 | 0.07 | -0.03 | 0.03 | -0.01 | 0.01 |
| ARG | 0.10 | 0.12 | -0.05 | 0.02 | 0.04 | -0.03 |
| ARM | 0.08 | 0.06 | -0.01 | 0.00 | 0.01 | -0.01 |
| ATG | 0.00 | -0.02 | 0.00 | 0.00 | 0.00 | 0.00 |
| **AUS** | **6.83** | **6.64** | **1.36** | **-0.35** | **-1.16** | **1.00** |
| AUT | 0.02 | 0.01 | -0.01 | 0.00 | 0.01 | -0.01 |
| AZE | -0.01 | -0.02 | 0.00 | 0.00 | 0.00 | 0.00 |
| BDI | 0.02 | -0.01 | -0.01 | 0.02 | -0.02 | 0.02 |
| BEN | 0.00 | -0.02 | 0.00 | 0.00 | -0.01 | 0.00 |

| | | | | | |
|------|------|------|------|------|------|
| BFA | 0.00 | -0.02 | -0.02 | 0.02 | 0.00 | 0.00 |
| BGD | 0.00 | 0.01 | -0.05 | 0.03 | 0.02 | -0.02 |
| BGR | 0.00 | -0.02 | 0.00 | 0.00 | 0.00 | 0.00 |
| BHR | 0.21 | 0.17 | -0.03 | 0.04 | -0.01 | 0.01 |
| BHS | 0.02 | 0.01 | -0.05 | 0.05 | 0.00 | 0.00 |
| BIH | 0.00 | -0.01 | 0.00 | 0.00 | 0.01 | -0.01 |
| BLZ | 0.00 | 0.01 | -0.08 | 0.07 | 0.01 | -0.01 |
| BMU | 0.01 | 0.01 | -0.02 | 0.01 | 0.01 | -0.01 |
| BOL | 0.65 | 0.56 | -0.09 | 0.17 | -0.09 | 0.08 |
| BRA | 0.24 | 0.17 | -0.01 | 0.00 | 0.00 | 0.00 |
| BRB | 0.00 | 0.08 | -0.23 | 0.16 | 0.08 | -0.07 |
| BRN | 0.36 | 0.32 | -0.02 | 0.05 | -0.04 | 0.04 |
| BTN | 0.00 | -0.02 | -0.03 | 0.03 | 0.00 | 0.00 |
| BWA | -0.01 | -0.03 | 0.00 | -0.01 | 0.00 | 0.00 |
| CAF | 0.00 | 0.00 | -0.02 | 0.01 | 0.01 | -0.01 |
| **CAN** | **3.19** | **2.85** | **1.03** | **-0.48** | **-0.63** | **0.55** |
| CCK | 0.00 | -0.02 | -0.55 | 0.56 | -0.01 | 0.01 |
| CHE | 0.07 | 0.06 | -0.02 | 0.01 | 0.02 | -0.01 |
| **CHL** | **4.05** | **3.71** | **0.99** | **-0.43** | **-0.64** | **0.56** |
| CHN | 0.32 | 0.23 | -0.01 | -0.02 | 0.03 | -0.03 |
| CIV | 0.00 | -0.02 | -0.02 | 0.02 | 0.00 | 0.00 |
| CMR | 0.00 | -0.02 | 0.00 | 0.00 | 0.00 | 0.00 |
| COG | 0.00 | -0.02 | 0.01 | 0.00 | -0.01 | 0.01 |
| COK | 0.00 | 0.04 | -0.50 | 0.46 | 0.05 | -0.04 |
| COL | 0.55 | 0.44 | -0.02 | 0.07 | -0.06 | 0.05 |
| COM | 0.00 | 0.03 | -0.05 | 0.01 | 0.04 | -0.04 |
| CPV | 0.00 | -0.01 | -0.01 | 0.00 | 0.00 | 0.00 |
| CRI | 0.17 | 0.12 | -0.02 | 0.07 | -0.05 | 0.04 |
| CUB | 0.03 | 0.05 | -0.10 | 0.07 | 0.04 | -0.03 |
| CXR | 0.00 | 0.17 | -0.54 | 0.38 | 0.19 | -0.16 |
| CYM | 0.00 | -0.01 | -0.05 | 0.04 | 0.01 | -0.01 |
| CYP | 0.15 | 0.11 | -0.01 | 0.00 | 0.01 | -0.01 |
| CZE | 0.01 | -0.01 | 0.00 | 0.00 | 0.01 | -0.01 |
| DEU* | 0.05 | 0.03 | -0.01 | 0.00 | 0.01 | -0.01 |
| DJI | 0.00 | -0.02 | 0.00 | 0.01 | -0.01 | 0.01 |
| DMA | 0.00 | -0.01 | -0.04 | 0.04 | -0.01 | 0.01 |
| DNK | 0.04 | 0.03 | -0.01 | 0.00 | 0.02 | -0.01 |
| DOM | 0.01 | -0.03 | -0.02 | 0.05 | -0.04 | 0.03 |
| DZA | -0.05 | -0.08 | 0.00 | 0.01 | -0.01 | 0.01 |
| ECU | 0.46 | 0.35 | -0.01 | 0.06 | -0.06 | 0.05 |
| EGY | 0.09 | 0.05 | 0.00 | 0.01 | -0.01 | 0.01 |
| ERI | 0.01 | -0.02 | 0.00 | 0.01 | -0.01 | 0.01 |
| ESH | 0.02 | 0.55 | -0.60 | 0.29 | 0.37 | -0.32 |
| ESP | 0.06 | 0.03 | -0.01 | 0.00 | 0.01 | -0.01 |
| EST | 0.01 | -0.01 | 0.00 | 0.00 | 0.01 | -0.01 |
| ETH | 0.15 | 0.11 | 0.00 | 0.00 | 0.00 | 0.00 |
| FIN | 0.05 | 0.03 | -0.01 | 0.00 | 0.01 | -0.01 |

| | | | | | |
|---|---|---|---|---|---|
| FJI | 0.91 | 0.77 | -0.12 | 0.20 | -0.10 | 0.08 |
| FLK | 0.00 | 0.00 | -0.01 | 0.00 | 0.01 | -0.01 |
| FRA | 0.04 | 0.02 | -0.01 | 0.00 | 0.01 | -0.01 |
| FRO | 0.00 | -0.01 | 0.00 | -0.01 | 0.01 | -0.01 |
| FSM | 0.00 | -0.05 | -0.07 | 0.12 | -0.06 | 0.05 |
| GAB | 0.00 | 0.00 | -0.03 | 0.01 | 0.02 | -0.01 |
| GBR | 0.11 | 0.08 | -0.02 | 0.02 | 0.00 | 0.00 |
| GEO | 0.11 | 0.09 | -0.02 | 0.00 | 0.02 | -0.02 |
| GHA | 0.00 | -0.02 | -0.01 | 0.02 | -0.01 | 0.01 |
| GIB | 0.00 | -0.02 | 0.00 | 0.00 | 0.00 | 0.00 |
| GIN | 0.00 | -0.02 | -0.01 | 0.01 | 0.00 | 0.00 |
| GMB | 0.00 | -0.01 | -0.01 | 0.01 | 0.00 | 0.00 |
| GNB | 0.00 | -0.02 | 0.00 | 0.01 | -0.01 | 0.01 |
| GNQ | 0.00 | -0.02 | 0.01 | 0.00 | -0.01 | 0.01 |
| GRC | 0.01 | -0.01 | 0.00 | 0.00 | 0.00 | 0.00 |
| GRD | 0.00 | -0.03 | -0.04 | 0.06 | -0.02 | 0.02 |
| GRL | 0.00 | 0.00 | -0.01 | 0.00 | 0.02 | -0.02 |
| GTM | 0.03 | 0.00 | -0.07 | 0.09 | -0.03 | 0.02 |
| GUY | 0.00 | 0.11 | -0.12 | 0.03 | 0.11 | -0.09 |
| HKG | 0.07 | 0.06 | -0.04 | 0.03 | 0.01 | -0.01 |
| HND | 0.00 | -0.02 | -0.03 | 0.05 | -0.02 | 0.02 |
| HRV | -0.01 | -0.03 | 0.00 | -0.01 | 0.01 | 0.00 |
| HTI | 0.00 | -0.04 | -0.02 | 0.06 | -0.04 | 0.03 |
| HUN | 0.01 | 0.00 | -0.01 | 0.00 | 0.01 | -0.01 |
| IDN | 0.44 | 0.40 | -0.05 | 0.02 | 0.03 | -0.03 |
| IND | 0.13 | 0.08 | -0.01 | 0.01 | -0.01 | 0.01 |
| IRL | 0.03 | 0.03 | -0.02 | 0.01 | 0.01 | -0.01 |
| IRN | 0.03 | 0.00 | 0.00 | 0.00 | 0.00 | 0.00 |
| IRQ | -0.03 | -0.06 | 0.00 | 0.01 | -0.01 | 0.01 |
| ISL | 0.04 | 0.02 | -0.01 | 0.01 | 0.00 | 0.00 |
| ISR | 0.05 | 0.02 | -0.01 | 0.01 | 0.00 | 0.00 |
| ITA | 0.05 | 0.03 | -0.01 | 0.00 | 0.01 | -0.01 |
| JAM | 0.02 | 0.04 | -0.07 | 0.05 | 0.03 | -0.02 |
| JOR | 0.04 | 0.00 | 0.00 | 0.02 | -0.02 | 0.01 |
| **JPN** | **3.64** | **3.50** | **1.47** | **-0.87** | **-0.69** | **0.59** |
| KAZ | 0.07 | 0.05 | -0.01 | -0.01 | 0.02 | -0.02 |
| KEN | 0.05 | 0.01 | 0.00 | 0.02 | -0.03 | 0.02 |
| KGZ | -0.06 | -0.08 | 0.01 | -0.02 | 0.01 | -0.01 |
| KHM | 0.09 | 0.10 | -0.11 | 0.09 | 0.01 | -0.01 |
| KIR | 0.00 | 0.18 | -0.48 | 0.33 | 0.18 | -0.15 |
| KNA | 0.00 | -0.02 | -0.03 | 0.04 | -0.02 | 0.02 |
| KOR | 0.33 | 0.29 | -0.03 | 0.02 | 0.02 | -0.02 |
| KWT | 0.36 | 0.33 | -0.04 | 0.03 | 0.02 | -0.01 |
| LAO | 0.00 | 0.10 | -0.08 | 0.01 | 0.08 | -0.07 |
| LBN | 0.00 | -0.02 | 0.00 | 0.01 | -0.01 | 0.01 |
| LBR | 0.00 | -0.01 | -0.06 | 0.06 | 0.01 | -0.01 |
| LBY | 0.00 | -0.02 | 0.00 | 0.00 | 0.00 | 0.00 |

| | | | | | |
|---|---|---|---|---|---|
| LCA | 0.00 | -0.05 | -0.01 | 0.05 | -0.05 | 0.04 |
| LKA | 0.14 | 0.10 | -0.01 | 0.04 | -0.03 | 0.02 |
| LSO | 0.00 | -0.04 | 0.03 | 0.00 | -0.03 | 0.03 |
| LTU | 0.00 | -0.01 | 0.00 | 0.00 | 0.01 | -0.01 |
| LVA | 0.01 | 0.00 | -0.01 | 0.00 | 0.01 | -0.01 |
| MAC | 0.01 | -0.02 | 0.00 | 0.01 | -0.01 | 0.01 |
| MAR | 0.00 | -0.02 | 0.00 | 0.00 | 0.00 | 0.00 |
| MDA | 0.01 | 0.00 | 0.00 | 0.00 | 0.01 | -0.01 |
| MDG | 0.01 | 0.01 | -0.02 | 0.01 | 0.02 | -0.01 |
| MDV | 0.12 | 0.07 | 0.00 | 0.03 | -0.03 | 0.03 |
| **MEX** | **2.69** | **2.39** | **0.99** | **-0.55** | **-0.51** | **0.44** |
| MHL | 0.00 | 0.02 | -0.07 | 0.05 | 0.03 | -0.02 |
| MKD | 0.00 | -0.02 | 0.00 | 0.00 | 0.00 | 0.00 |
| MLI | 0.00 | -0.03 | 0.00 | 0.01 | -0.01 | 0.01 |
| MLT | 0.01 | 0.00 | -0.02 | 0.01 | 0.00 | 0.00 |
| MMR | 0.33 | 0.28 | -0.01 | 0.00 | 0.01 | -0.01 |
| MNG | 0.31 | 0.28 | -0.03 | 0.01 | 0.03 | -0.03 |
| MNP | 0.00 | 0.17 | -0.28 | 0.13 | 0.18 | -0.16 |
| MOZ | 0.00 | -0.02 | -0.01 | 0.02 | -0.01 | 0.00 |
| MRT | 0.00 | 0.12 | -0.13 | 0.01 | 0.13 | -0.11 |
| MSR | 0.00 | 0.05 | -0.07 | 0.02 | 0.06 | -0.05 |
| MUS | 0.05 | 0.03 | -0.03 | 0.04 | -0.01 | 0.01 |
| MWI | 0.13 | 0.10 | -0.02 | 0.00 | 0.02 | -0.02 |
| **MYS** | **2.62** | **2.35** | **0.82** | **-0.44** | **-0.43** | **0.37** |
| NAM | 0.01 | -0.01 | 0.00 | -0.01 | 0.01 | -0.01 |
| NCL | 0.00 | 0.09 | -0.20 | 0.11 | 0.10 | -0.09 |
| NER | 0.03 | 0.01 | -0.01 | 0.01 | 0.00 | 0.00 |
| NFK | 0.00 | -0.01 | -0.44 | 0.44 | 0.00 | 0.00 |
| NGA | -0.01 | -0.05 | 0.00 | 0.01 | -0.01 | 0.01 |
| NIC | 0.00 | 0.01 | -0.10 | 0.09 | 0.01 | -0.01 |
| NIU | 0.00 | -0.02 | -0.29 | 0.30 | 0.00 | 0.00 |
| NLD | 0.02 | 0.02 | -0.03 | 0.02 | 0.01 | -0.01 |
| NOR | 0.09 | 0.06 | -0.01 | 0.01 | 0.01 | 0.00 |
| NPL | 0.00 | -0.01 | -0.01 | 0.01 | 0.00 | 0.00 |
| NRU | 0.00 | 0.15 | -0.55 | 0.43 | 0.15 | -0.13 |
| **NZL** | **4.12** | **3.74** | **1.22** | **-0.53** | **-0.79** | **0.69** |
| OMN | 0.24 | 0.20 | -0.03 | 0.07 | -0.05 | 0.04 |
| PAK | 0.00 | -0.01 | -0.04 | 0.04 | 0.00 | 0.00 |
| PAN | 0.26 | 0.21 | -0.09 | 0.14 | -0.06 | 0.05 |
| PCN | 0.00 | -0.01 | -0.16 | 0.17 | -0.01 | 0.01 |
| **PER** | **5.80** | **5.65** | **1.42** | **-0.51** | **-1.04** | **0.90** |
| PHL | 0.59 | 0.57 | -0.08 | 0.05 | 0.04 | -0.03 |
| PLW | 0.00 | -0.02 | -0.08 | 0.10 | -0.02 | 0.02 |
| PNG | 0.06 | 0.26 | -0.42 | 0.33 | 0.11 | -0.09 |
| POL | 0.01 | -0.01 | 0.00 | 0.00 | 0.01 | -0.01 |
| PRK | 0.00 | 0.04 | -0.03 | -0.01 | 0.05 | -0.04 |
| PRT | 0.01 | -0.01 | 0.00 | 0.00 | 0.01 | -0.01 |

| | | | | | |
|---|---|---|---|---|---|
| PRY | 0.23 | 0.22 | -0.05 | 0.02 | 0.04 | -0.03 |
| PYF | 0.00 | 0.01 | -0.11 | 0.10 | 0.01 | -0.01 |
| QAT | 0.43 | 0.39 | -0.04 | 0.02 | 0.02 | -0.01 |
| ROM | 0.00 | -0.02 | 0.00 | 0.00 | 0.01 | -0.01 |
| RUS | 0.07 | 0.05 | -0.01 | 0.00 | 0.01 | -0.01 |
| RWA | 0.00 | 0.00 | -0.03 | 0.02 | 0.01 | -0.01 |
| SAU | 0.23 | 0.20 | -0.03 | 0.03 | -0.01 | 0.01 |
| SEN | 0.00 | -0.02 | 0.00 | 0.01 | -0.01 | 0.01 |
| SGP | 0.02 | 0.10 | -0.18 | 0.10 | 0.09 | -0.08 |
| SHN | 0.00 | 0.01 | -0.01 | 0.00 | 0.02 | -0.01 |
| SLB | 0.00 | 0.03 | -0.28 | 0.26 | 0.03 | -0.02 |
| SLE | 0.00 | 0.00 | -0.04 | 0.03 | 0.01 | -0.01 |
| SLV | 0.00 | -0.05 | -0.02 | 0.07 | -0.05 | 0.05 |
| SMR | 0.00 | 0.02 | -0.02 | -0.01 | 0.03 | -0.03 |
| SOM | 0.00 | -0.03 | 0.00 | 0.02 | -0.02 | 0.02 |
| SPM | 0.00 | -0.02 | -0.14 | 0.15 | -0.01 | 0.01 |
| STP | 0.00 | -0.01 | -0.03 | 0.03 | 0.00 | 0.00 |
| SUR | 0.02 | 0.05 | -0.06 | 0.04 | 0.03 | -0.03 |
| SVK | 0.00 | -0.01 | 0.00 | 0.00 | 0.01 | -0.01 |
| SVN | 0.00 | -0.01 | 0.00 | 0.00 | 0.01 | -0.01 |
| SWE | 0.04 | 0.03 | -0.01 | 0.00 | 0.01 | -0.01 |
| SWZ | 0.00 | 0.01 | 0.00 | -0.01 | 0.01 | -0.01 |
| SYC | 0.00 | 0.00 | -0.02 | 0.01 | 0.01 | -0.01 |
| SYR | -0.01 | -0.04 | 0.00 | 0.00 | 0.00 | 0.00 |
| TCA | 0.00 | 0.02 | -0.05 | 0.03 | 0.02 | -0.02 |
| TCD | 0.00 | -0.04 | 0.03 | 0.00 | -0.04 | 0.03 |
| TGO | 0.00 | -0.02 | 0.00 | 0.00 | -0.01 | 0.01 |
| THA | 0.26 | 0.30 | -0.15 | 0.09 | 0.07 | -0.06 |
| TJK | -0.03 | -0.05 | 0.00 | -0.01 | 0.01 | -0.01 |
| TKL | 0.00 | 0.03 | -0.02 | -0.01 | 0.04 | -0.03 |
| TKM | 0.00 | -0.02 | 0.00 | 0.00 | 0.00 | 0.00 |
| TMP | 0.00 | 0.05 | -0.12 | 0.07 | 0.05 | -0.04 |
| TON | 0.21 | 0.20 | -0.10 | 0.13 | -0.03 | 0.02 |
| TTO | 0.04 | 0.01 | -0.02 | 0.05 | -0.03 | 0.03 |
| TUN | -0.01 | -0.02 | 0.00 | 0.00 | 0.01 | 0.00 |
| TUR | 0.02 | -0.01 | 0.00 | 0.00 | 0.00 | 0.00 |
| TUV | 0.00 | -0.04 | -0.26 | 0.31 | -0.06 | 0.05 |
| TZA | 0.04 | 0.02 | -0.01 | 0.01 | -0.01 | 0.01 |
| UGA | 0.00 | -0.02 | -0.02 | 0.03 | -0.01 | 0.01 |
| UKR | 0.01 | -0.01 | 0.00 | 0.00 | 0.01 | -0.01 |
| URY | 0.14 | 0.12 | -0.02 | 0.01 | 0.02 | -0.02 |
| USA | 1.43 | 1.27 | -0.05 | 0.07 | -0.02 | 0.02 |
| UZB | 0.00 | 0.01 | -0.02 | 0.00 | 0.03 | -0.02 |
| VCT | 0.00 | -0.02 | -0.02 | 0.02 | -0.01 | 0.01 |
| VEN | 0.02 | 0.00 | -0.05 | 0.06 | -0.02 | 0.01 |
| VGB | 0.00 | -0.01 | 0.00 | 0.00 | 0.00 | 0.00 |
| **VNM** | **3.88** | **3.56** | **1.09** | **-0.53** | **-0.65** | **0.56** |

| | | | | | |
|-----|------|------|-------|------|------|------|
| VUT | 0.00 | 0.01 | -0.16 | 0.15 | 0.02 | -0.02 |
| WLF | 0.00 | 0.02 | -0.21 | 0.19 | 0.03 | -0.02 |
| WSM | 0.00 | 0.23 | -0.34 | 0.25 | 0.11 | -0.09 |
| YEM | 0.10 | 0.06 | 0.00 | 0.02 | -0.02 | 0.01 |
| ZAF | 0.25 | 0.21 | -0.02 | 0.01 | 0.01 | -0.01 |
| ZAR | 0.00 | -0.01 | -0.01 | 0.00 | 0.01 | -0.01 |
| ZMB | 0.00 | 0.00 | -0.02 | 0.01 | 0.01 | -0.01 |
| ZWE | 0.00 | -0.01 | 0.00 | -0.01 | 0.02 | -0.01 |

# 6   Codes

All the codes for the simulation are available at

[https://github.com/pachadotdev/general-equilibrium-ftas-chile](https://github.com/pachadotdev/general-equilibrium-ftas-chile).